\documentclass[twocolumn]{aastex7}
\usepackage{soul}
\usepackage{comment}
\usepackage{xcolor}

\newcommand{\revise}[1]{#1}

\begin{document}


\title{A tale of dynamical instabilities and giant impacts in the radius valley}
\author[sname='Shibata']{Sho Shibata}
\affiliation{Department of Earth, Environmental and Planetary Sciences, MS 126,  Rice University, Houston, TX 77005, USA}
\email[show]{s.shibata423@gmail.com}

\author[sname='Izidoro']{Andre Izidoro} 
\affiliation{Department of Earth, Environmental and Planetary Sciences, MS 126,  Rice University, Houston, TX 77005, USA}
\email{}



\begin{abstract}
The size distribution of planets with radii between 1 and 4~$R_{\oplus}$ peaks near 1.4 and 2.2~$R_{\oplus}$, with a dip around 1.8~$R_{\oplus}$ --the so-called ``radius valley.'' Recent statistical analyses suggest that planets within this valley ($1.5 < R < 2$~$R_{\oplus}$) tend to have slightly higher orbital eccentricities than those outside it. The origin of this dynamical signature remains unclear.\revise{ We revisit the ``breaking the chains'' formation model and propose that late dynamical instabilities --occurring after disk dispersal-- may account for the elevated eccentricities observed in the radius valley. Our simulations show that sub-valley planets ($R < 2R_{\oplus}$) are generally rocky, while those beyond the valley ($R > 2R_{\oplus}$) are typically water-rich. Rocky planets that undergo strong dynamical instabilities and numerous late giant impacts have their orbits excited and their radii increased, ultimately placing them into the radius valley. In contrast, the larger, water-rich planets just beyond the valley experience weaker instabilities and fewer impacts, resulting in lower eccentricities. This contrast leads to a peak in the eccentricity distribution within the valley.} The extent to which planets in the radius valley are dynamically excited depends sensitively on \revise{the orbital architecture before the orbital instability.} Elevated eccentricities among radius valley planets arise primarily in scenarios that form a sufficiently large number of rocky planets within 100 days (typically $\gtrsim 5$) prior to instability, and that also \revise{host} external perturbers ($P > 100$ days), which further amplify the strength of dynamical instabilities. 
\end{abstract}

\section{Introduction} 

NASA's Kepler mission revealed over 5,000 exoplanets and constrained various physical features of these planets. Super-Earths and mini-Neptunes, planets of sizes ranging from 1 to 4 Earth radii, are the most common type of planets observed today. Their size distribution is bimodal, with peaks at approximately 1.4 and 2.4 Earth radii and a gap around 1.8 Earth radii, known as the ``exoplanet radius valley'' \citep{Fulton+2017, Fulton+2018}. This feature suggests a composition dichotomy: planets of about 1.4 Earth radii are likely dominated by silicate compositions, while those with sizes around 2.4 Earth radii may have significant volatile content in their cores or primordial H-He rich atmospheres~\citep{Kuchner03, Fortney+2007, Adams+2008, Rogers+2010, Owen+2017, Zeng+2019, Venturini+2020, Izidoro+2022, Burn+2024, Shibata+2025, Nielsen+2025}. While many formation models broadly succeeded in accounting for this dichotomy in the planet size distribution, the origins of these planets still remain a subject of strong debate \citep{Owen+2017, Gupta+20,
Venturini+2020, Izidoro+2022, Burn+2024, Shibata+2025, Nielsen+2025}.

\begin{figure}[ht!]
    \plotone{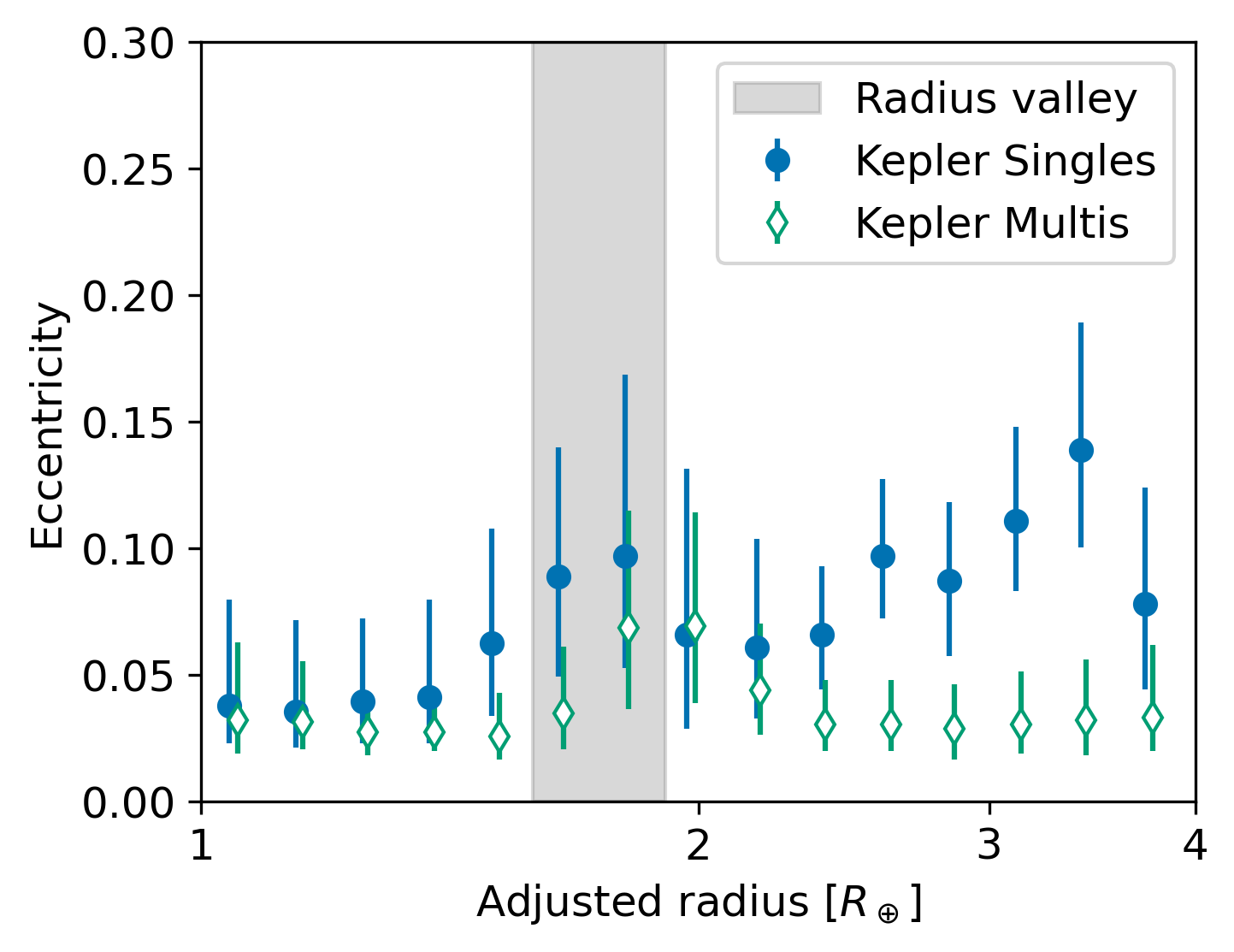}
    \caption{
    Eccentricity distribution of close-in super-Earths and mini-Neptunes. The data is extracted from Fig. 4 in \citet{Gilbert+2025} through visual inspection using PlotDigitizer ({\rm https://plotdigitizer.com/app}). The blue points show the mean eccentricity for Kepler planets observed as single planets, and the green points for those observed as multiple planets. 
    }
    \label{fig: Rad_Ecc_Obs}
\end{figure}

Recently, \citet{Gilbert+2025} identified a new dynamical feature related to the radius valley. By grouping planets based on properties such as radius and orbital period, their statistical analysis points to an interesting trend: planets located within the radius valley (around 1.8 Earth radii) exhibit elevated mean eccentricities compared to those on either side.  Figure~\ref{fig: Rad_Ecc_Obs} shows the observed eccentricity distribution of planets with the orbital periods P$<100$ days reported in their work. In our work, we focus on planets within the radius valley (grey region in Figure~\ref{fig: Rad_Ecc_Obs}).

The radius valley itself has been attributed to a range of mechanisms, including photo-evaporation \citep{Lopez+2013, Owen+2017}, core-powered mass-loss \citep{Ginzburg+2018, Gupta+20}, impact-induced mass loss \citep{Liu+2015, Inamdar+2016, Biersteker+2019, Kegerreis+2020, Matsumoto+2021, Chance+2022}, stripping the primordial atmospheres of close-in planets, or to a dichotomy in their bulk compositions \citep{Venturini+2020, Izidoro+2022, Burn+2024, Shibata+2025}. In contrast, the eccentricity trend for radius valley planets is likely driven by a dynamical process, but its true origin remains elusive. Taken together, these two observational features impose a powerful constraint on formation models.

This work represents the first attempt to understand the origins of this correlation within a dynamical model of planet formation. We revisit recent simulations of the breaking the chains model, modeling the formation of super-Earths and mini-Neptunes presented in \citet{Shibata+2025}. These simulations broadly reproduce a number of observational constraints as the period ratio, the planet size-ratio, planet multiplicity, and size distributions of these planets, including the radius valley. Given this broad math to observations, these simulations provide an ideal sample to investigate the origin of the eccentricity trend identified by \citet{Gilbert+2025}. Here, we show that the trend reported by \citet{Gilbert+2025} is consistent with late dynamical instabilities and giant impacts among rocky planets, leading to the formation of a dynamically excited population of planets within the radius valley.
 
This paper is structured as follows. In Section~\ref{sec: simulation_data}, we describe the simulation setup from \citet{Shibata+2025}. In Section~\ref{sec: Results}, we present our main results. In Section~\ref{sec: discussion}, we discuss our results by comparing them with observational data. Finally, Section~\ref{sec: conclusion} provides a summary and concluding remarks.

\section{Simulation data}\label{sec: simulation_data}

In the simulations of \citet{Shibata+2025}, super-Earths and mini-Neptunes form from two rings of planetesimals, as recently invoked in recent solar system formation models~\citep[e.g.][]{Izidoro+2021}. The inner and outer disks are located from 0.5-1.5 au and from 8-15 au, respectively. Simulations account for both planetesimal and pebble accretion processes.

The initial total mass of the inner planetesimal ring $M_\mathrm{disk, in}$ and the gaseous disk's lifetime $\tau_\mathrm{disk, life}$ are input parameters of our simulations. We select a sub-sample of the simulations of \citet{Shibata+2025} with the specific parameters listed in Tab.~\ref{tab: simulations}. The {\it parameter\_ID} given by the label {\it MxTy} indicates that the simulation assumes an inner planetesimal ring with mass $M_\mathrm{disk, in} = xM_\oplus$ and a disk dispersal timescale of $\tau_\mathrm{disk, life} = y$ Myr. For each parameter set, we have 50 simulations considering slightly different initial distributions of planetary seeds. Therefore, we analyze 200 simulations in total in this work. In our analysis, we focus on planets with masses greater than $0.1 M_\oplus$ and orbital periods shorter than 100 days. 
We use the term ``protoplanets'' to refer to objects existing during the simulation and ``planets'' to refer to those objects that survive at the end of the simulation and have masses larger than 0.1$M_{\oplus}$.

Planetesimals in the inner ring mainly grow via planetesimal accretion and typically become rocky protoplanets~\citep{Shibata+2025}. On the other hand, planetesimals in the outer ring grow mainly via pebble accretion and generally become icy protoplanets~\citep{Shibata+2025}. As these simulations account for gas-driven planet migration, rocky and icy protoplanets growing in these rings eventually migrate inwards and enter the inner disk (e.g $P<$ 100 days). If protoplanets migrate sufficiently inward during disk dispersal -- a process that may depend on their formation location and the timescale of the disk dispersal -- they can be trapped at the disk’s inner edge near 0.1~au, forming an extended resonant chain of planets captured in mean-motion resonances.

The simulations of~\cite{Shibata+2025} integrate the evolution of planetary systems up to 50 Myrs. After disk dissipation, a large fraction of the planetary systems undergo orbital instabilities. During orbital instabilities, planets experience late giant impacts. We define giant impacts as those where the mass ratio between the impactor and target is higher than 0.1. Our simulations assume that late giant impact triggers atmospheric mass loss as suggested in \citet{Biersteker+2019}. Collisions are considered perfect merging events that conserve mass and linear momentum (we refer the reader to \cite{estevesetal22} for a discussion on this assumption). We calculate the planetary radius of each planet --depending on its specific rocky or water-rich composition, which is an output of the simulations-- using the interior models by \citet{Zeng+2019}. We also account for the effects of atmospheric mass loss due to the photoevaporation \citep{Lopez+2013, Owen+2019}.

\begin{table}
    \centering
    \begin{tabular}{|c|c|c|c|}
        \hline
        parameter\_ID & $M_\mathrm{disk, in}$ & $\tau_\mathrm{disk, life}$ 
        \\
        \hline
        {\it M3T2} & $3 M_\oplus$ & $2$ Myrs \\
        {\it M3T3} & $3 M_\oplus$ & $3$ Myrs \\
        {\it M6T2} & $6 M_\oplus$ & $2$ Myrs \\
        {\it M6T3} & $6 M_\oplus$ & $3$ Myrs \\
        \hline
    \end{tabular}
    \caption{
    Simulation data and parameters used in this study. 
    }
    \label{tab: simulations}
\end{table}

\begin{figure}[ht!]
    \plotone{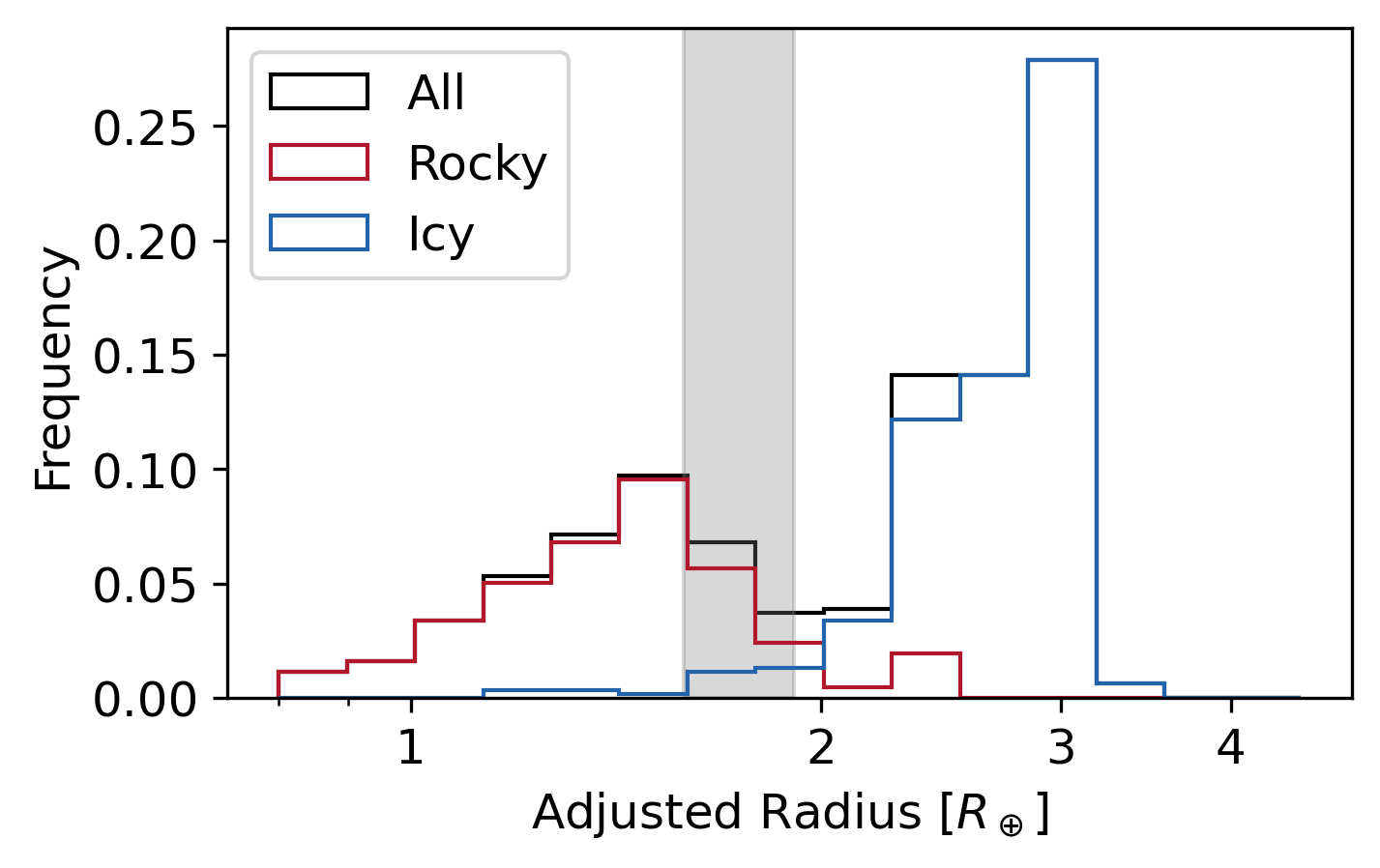}
    \caption{
    Planet adjusted radius distribution of \citet{Shibata+2025} combining different sets of simulations. The red and blue lines show the distribution of rocky and icy planets, respectively. The black line shows the distribution of all planets. The gray area is the radius gap ($1.59 R_\oplus < R_\mathrm{p, adj} < 1.90 R_\oplus$). We use the adjusted radius to account for the shift of the radius valley along the orbital period. Note that we show the raw data, and the observational bias is not included.
    }
    \label{fig: R_hist}
\end{figure}

Figure~\ref{fig: R_hist} shows the radius distribution of planets used in this study. It shows the raw data of our simulations without applying observational bias. Note that we plot  the ``adjusted radius'' introduced in \citet{Ho+2023} to account for the shift of the radius valley along the orbital period, which is given by
\begin{equation}
    R_\mathrm{p,adj} = R_\mathrm{p}  \left( \frac{P}{10~\mathrm{day}} \right)^{m}, \label{eq: Rpadj}
\end{equation}
where $R_\mathrm{p}$ is the planet's radius and $m$ is set to $0.10$ \citep{Petigura+2022}. The red and blue lines show the rocky and icy planets. We define rocky planets as those with a water mass fraction of less than $10\%$. The gray area in Figure~\ref{fig: R_hist} marks the region where \citet{Gilbert+2025} observed an enhancement in eccentricity, corresponding to $1.6R_\oplus < R_\mathrm{p, adj} < 1.9R_\oplus$. In this paper, we refer to this region as the radius valley. In our model, rocky planets form below $\sim 2R_\oplus$ and icy planets above this threshold, creating two distinct populations. The radius valley lies between them. Over 80\% of the planets within the radius valley are rocky, formed from the inner planetesimal ring via planetesimal accretion~\citep{Shibata+2025}.

Note that the size distribution in Figure~\ref{fig: R_hist} differs from the nominal best fit solutions of \citet{Shibata+2025}, where observational biases are included and simulations from different scenarios are combined in varying proportions to match the observed distributions better. Figure~\ref{fig: R_hist} shows the results of all scenarios described in Table~\ref{tab: simulations}.

\section{Results: Dynamical Origins of Radius Valley Planets}\label{sec: Results}

\subsection{Examples of planetary instabilities}
\begin{figure}[ht!]
    \plotone{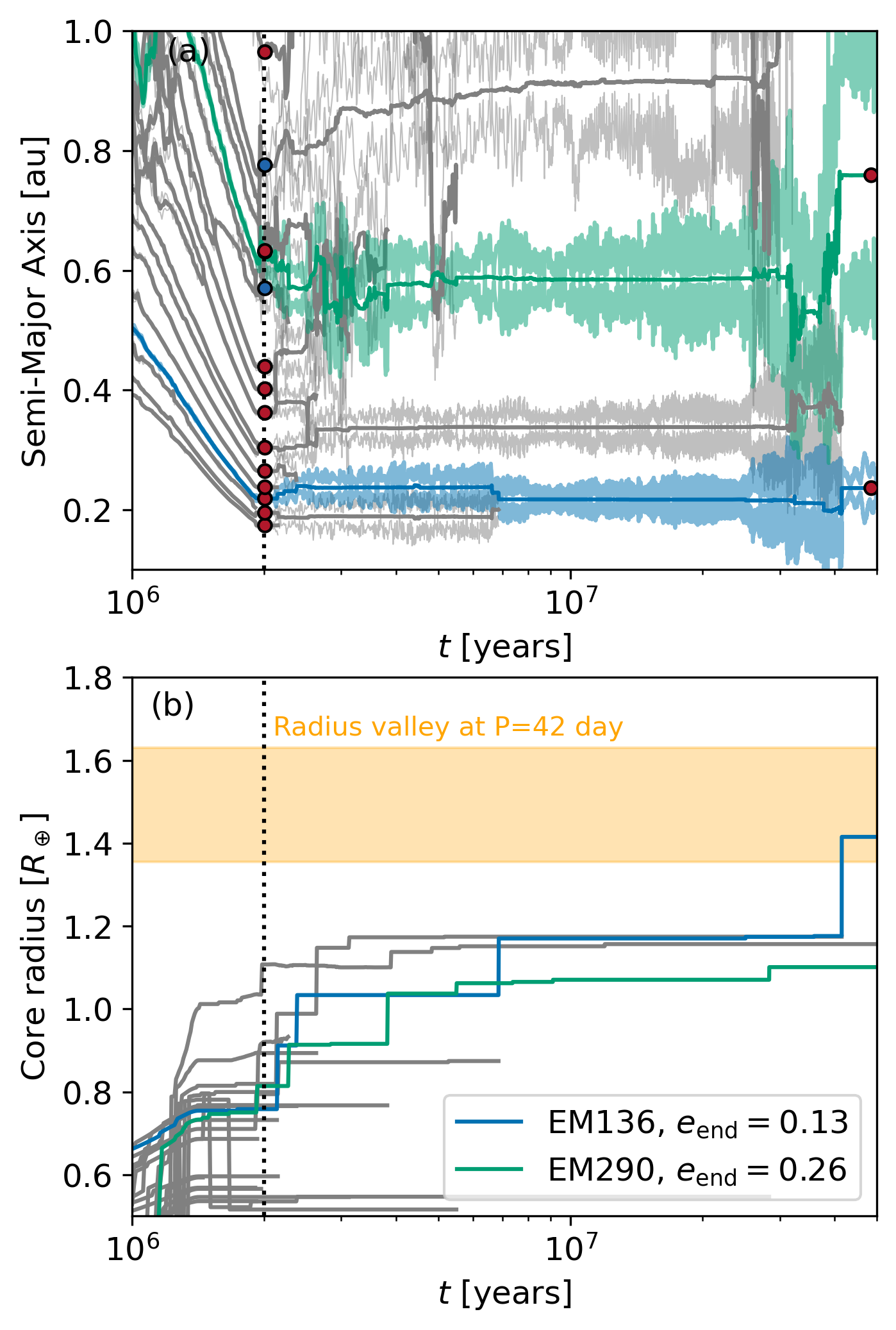}
    \caption{
    An example of the evolution of planets in the "breaking-the-chain" model. Here, we show one of the simulations obtained in {\it M3T2\_\it6}. Panel (a) shows the time evolution of the semi-major axis of planets. We plot protoplanets that grow larger than $0.1 M_\oplus$. The colored lines show the planets that survive at $t=50$ Myrs. The thin lines show the pericenter and apocenter of the planets. The vertical dotted black line is the time when the protoplanetary disk dissipates. The red and blue plots show that the protoplanet's composition is rocky and icy, respectively.
    Panel (b) shows the time evolution of the planetary core's radius. The orange area shows the radius valley at $P=53$ days, the final orbital period of {\it EM70}. We show the eccentricity at the end of the simulation $e_\mathrm{end}$ in the legend box. 
    }
    \label{fig: orbits_evol}
\end{figure}

To illustrate how planets populating the radius valley form and evolve, we selected a representative system of our {\it M3T2} simulation that produces a relatively high eccentricity planet inside the radius valley. This system is produced in our simulation {\it M3T2\_6}, where ``6'' represents the system's id (see Appendix~\ref{app: appendix2}). Panel (a) in  Fig.~\ref{fig: orbits_evol} shows the time evolution of the semi-major axes of protoplanets. Panel (b) illustrates the temporal evolution of protoplanet core sizes. Core sizes are calculated using mass-radius relationship fits of \cite{Zeng+2019}.

In our model, protoplanets grow mainly via planetesimal accretion inside the snowline and via pebble accretion outside the snowline. By the time of the disk dispersal at $t=2$ Myrs, rocky protoplanets growing in the ring around 1 au have migrated inward and reached closer-in orbits. In this simulation, 14 protoplanets enter the region inside $1$ au, and all inner nine protoplanets are rocky (see color-coded dots on Panel (a) in  Fig.~\ref{fig: orbits_evol} shown at the time of disk dispersal and at the end of the simulation; red represent rocky planets and blue represents water-rich planets). The masses of these rocky protoplanets range from $0.2 M_\oplus$ to $0.5 M_\oplus$, and all have core radii smaller than the radius gap ($R<1.59R_{\oplus}$). The outer four protoplanets have rocky and water-rich compositions. Water-rich planets are formed when they grow and migrate inward from the second ring or when the snowline moves inward (as the disk cools down) and sweeps the inner ring, allowing protoplanets in the inner ring to grow via pebble accretion \citep{Shibata+2025}.

Once the gaseous protoplanetary disk dissipates (dotted vertical line in Fig.~\ref{fig: orbits_evol}a-b), close-in protoplanets tend to enter a dynamical instability phase. Due to their mutual gravitational interactions, eccentricities gradually increase until orbits start to cross and protoplanets collide.  In  Fig.~\ref{fig: orbits_evol} we show in green and blue the evolution of protoplanets {\it EM136} and {\it EM290}, respectively, which are of particular interest given they are the two surviving planets in the inner regions in this simulation.

These two planets originated from the ring around 1 au and consequently have rocky compositions (water-mass fraction lower than 0.1). After the gas disk dispersal, {\it EM136} has undergone four late giant impacts in about 40 Myr, as evidenced by the stepwise increases in its size shown in Fig.~\ref{fig: orbits_evol}(b). We recall that we assume that giant impacts after disk dispersal strip planetary atmospheres~\citep{Shibata+2025}. Yet,  their core radius are updated as their masses evolve. In the case of {\it EM136}, the increase in mass due to four giant impacts eventually places this planet inside the radius valley (orange region in  Figure~\ref{fig: orbits_evol}(b) representing the radius valley for a planet at $P=42$~days; specifically {\it EM136})

In the simulation of Figure~\ref{fig: orbits_evol}, the dynamical instability is violent and long-lasting due to the relatively large number of protoplanets in the inner disk at the time of the disk dispersal. After multiple impacts, the final orbital eccentricity of {\it EM136} is approximately 0.1 at the end of the simulation. Although dynamical instabilities generally excite planetary eccentricities, giant impacts can counteract this process by damping dynamical excitation \citep{Matsumoto+2015}. Indeed, the apocenter and pericenter excursions of {\it EM136} show that this planet exhibited higher eccentricities between 20 and 40 Myr than at its final state.

Planet {\it EM290} also experiences multiple impacts, but its mass remains relatively small compared to {\it EM136}, and it does not ``enter'' into the radius valley. In contrast, its final orbital eccentricity is larger than that of  {\it EM136}, about 0.26, due to its relatively low mass and gravitational interactions with multiple protoplanets during the instability phase. After dynamical instabilities  and orbital crossing cease, planets dynamically evolve due to secular and resonant perturbations. Planet {\it EM136} represents a typical example of dynamical evolution for a planet that enters the radius valley and maintains a relatively high orbital eccentricity.

\begin{figure}[ht!]
    \plotone{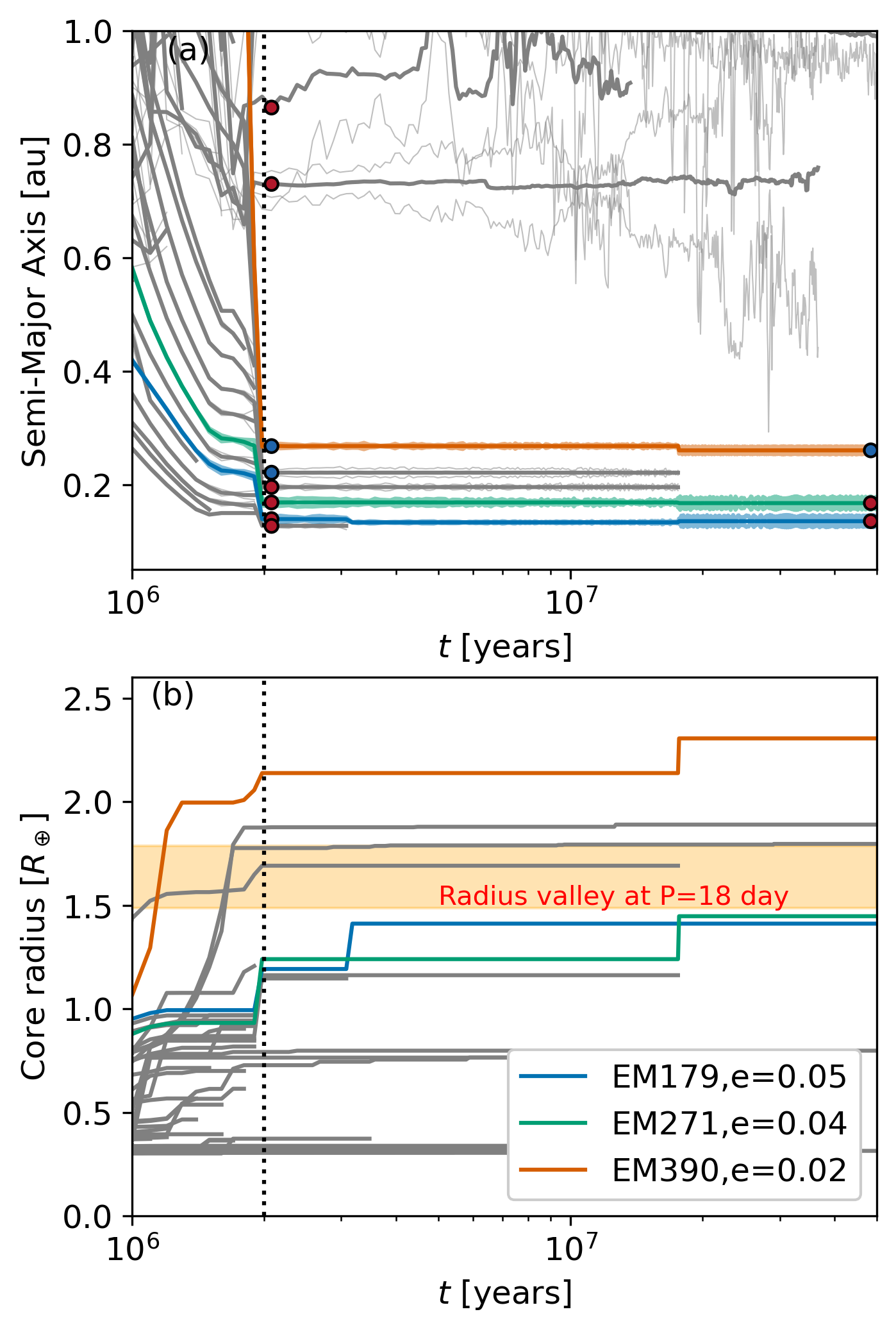}
    \caption{
    Same as Fig.~\ref{fig: orbits_evol}, but showing the simulations obtained in {\it M6T2\_\it45}.
    }
    \label{fig: orbits_evol_M6T2}
\end{figure}

Figure~\ref{fig: orbits_evol_M6T2} shows a planetary system from our {\it M6T2} scenario (system's id ``45''; see Appendix \ref{app: appendix2}), which exhibits several notable differences compared to the system shown in Figure~\ref{fig: orbits_evol}. The most striking differences-- arguably the most important, as we will discuss later-- are the number of protoplanets in the inner disk at the end of the gas disk phase. While the system in Figure~\ref{fig: orbits_evol} contains nine planets within 0.5 au, the system in Figure~\ref{fig: orbits_evol_M6T2} hosts only five. The post-disk dynamical instability in this case is relatively mild, with most planets undergoing only a single collision. Notably, no planets enter and survive within the radius valley. The only protoplanet that temporarily occupies this region is eventually accreted by an ice-rich body at around 20 Myr.

A key factor explaining the lower number of protoplanets at disk dispersal is the early inward migration of a water-rich protoplanet that formed beyond the snowline. As it migrated inward, it shepherded rocky bodies toward the star, eventually triggering collisions among them before disk dispersal. This process reduced the overall number of both rocky and water-rich protoplanets in the inner system. Also,  water-rich protoplanets -- by pushing the rocky protoplanets inwards -- separate/decouple the rocky protoplanet from the outer protoplanets. Different from {\it M3T2\_6} in Figure~\ref{fig: orbits_evol}, where nearly all protoplanets inside 1au  are gravitationally involved in the instability, the orbital instability in Figure \ref{fig: orbits_evol_M6T2} is largely confined to the inner five protoplanets and virtually decoupled from the outer eccentric protoplanets (some even beyond 1~au). Due to the mild dynamical instability and low number of giant impacts after disk dispersal, inner planets in Figure~\ref{fig: orbits_evol_M6T2} do not enter the radius valley nor show relatively larger orbital eccentricities.

In this section, we present two contrasting examples of dynamical instability to illustrate that the architecture of a planetary system at the time of disk dispersal -- particularly the number of planets in the inner regions-- plays a key role in shaping its final configuration. In the next section, we conduct a more comprehensive analysis using the full suite of simulated systems to assess whether radius valley planets in our simulations exhibit higher orbital eccentricities than those outside the valley in our different formation scenarios.

\subsection{Correlation Between Planetary Radius and Orbital Eccentricity}

\begin{figure}[ht!]
    \plotone{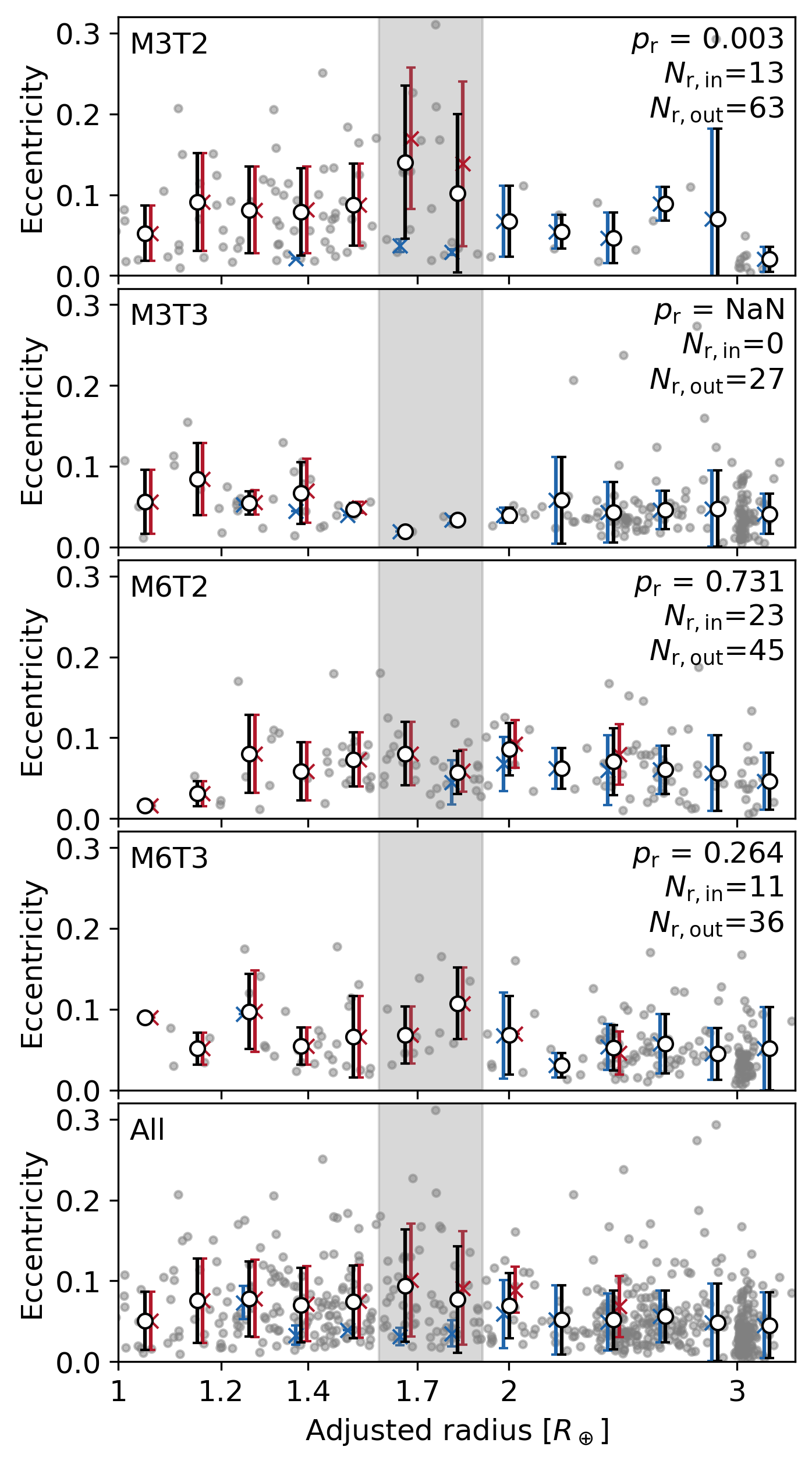}
    \caption{
    Eccentricity of planets as a function of their adjusted planetary radius. The gray plots show the raw data \revise{(individual planets)} obtained in our simulations, and the white plots show the mean eccentricity of the planets. 
    The grid size is the same as \citet{Gilbert+2025}, where 15 bins are taken between $1R_\oplus$ and $4R_\oplus$ in logarithmic scale. The error bar shows the standard deviation $1 \sigma$. 
    \revise{We also plot the mean eccentricity of rocky and icy planets with red and blue points.}    
    The gray area is the radius gap. The top four panels show the results obtained in each parameter set, and the bottom panel shows the results combining all simulations.$p_\mathrm{r}$ represents the p-value of KS-tests comparing the eccentricity distribution of rocky planets inside and outside the radius gap. We also show the number of rocky planets inside $N_\mathrm{r, in}$ and outside $N_\mathrm{r, out}$ the radius gap.
    }
    \label{fig: Mean_Ecc}
\end{figure}

In this section, we analyze the final eccentricity distribution of planets produced in our simulations.  We start our analysis by recalling that after the main giant impact phase, secular perturbations from neighboring planets can induce long-term variations in eccentricity (see Figure \ref{fig: orbits_evol}). To mitigate the effects of these fluctuations in our analysis of final eccentricities, we compute the time-averaged eccentricity of each planet over the final 5 Myr of the simulation. We have verified that our main results remain qualitatively unchanged when using a 1 Myr or 10 Myr averaging window instead.

Figure~\ref{fig: Mean_Ecc} shows the time-averaged eccentricity of planets as a function of their adjusted radius. Each gray dot represents the mean eccentricity of an individual planet, averaged over the final 5 Myr of its evolution, as described above. Open black circles indicate the mean eccentricities of planets within bins of adjusted radius (including rocky and water-rich planets). The bins are calculated using logarithmic spacing with 15 bins from 1 to 4 $R_{\oplus}$, following \citet{Gilbert+2025}. The black error bars represent the 1-$\sigma$ variation of planetary eccentricities within each bin. The mean eccentricities of rocky and water-rich planets are shown separately with red and blue error bars (and crosses), respectively. From top to bottom, the figure displays the results for each individual scenario, with the bottom panel showing the combined distribution from all systems. 


In the {\it M3T2} scenario, planets inside the radius valley (highlighted by the gray region) exhibit relatively higher orbital eccentricities than those outside the valley.
In contrast, the {\it M3T3} scenario produces virtually no planets within the radius valley. The {\it M6T2} and {\it M6T3} scenarios both produce planets inside the radius valley, but do not show any clear enhancement in orbital eccentricities relative to planets outside the valley. 
\revise{We defer a detailed explanation of the underlying mechanisms causing the observed trends and why the eccentricity peaks in the radius valley in some cases but not in others to the following sections.}

To perform a more systematic analysis, we apply a Kolmogorov–Smirnov (KS) test to compare the orbital eccentricity distributions of rocky planets inside and outside the radius valley. P-values and the number of planets inside and outside the radius valley are reported in Figure~\ref{fig: Mean_Ecc}. We find that only the {\it M3T2} scenario yields statistically distinct populations, with a p-value of about 0.003. In the {\it M6T2} and {\it M6T3} scenarios, although there appears to be a slight trend toward higher eccentricities for planets within the radius valley, the differences are not statistically significant.

\begin{figure}[ht!]
    \plotone{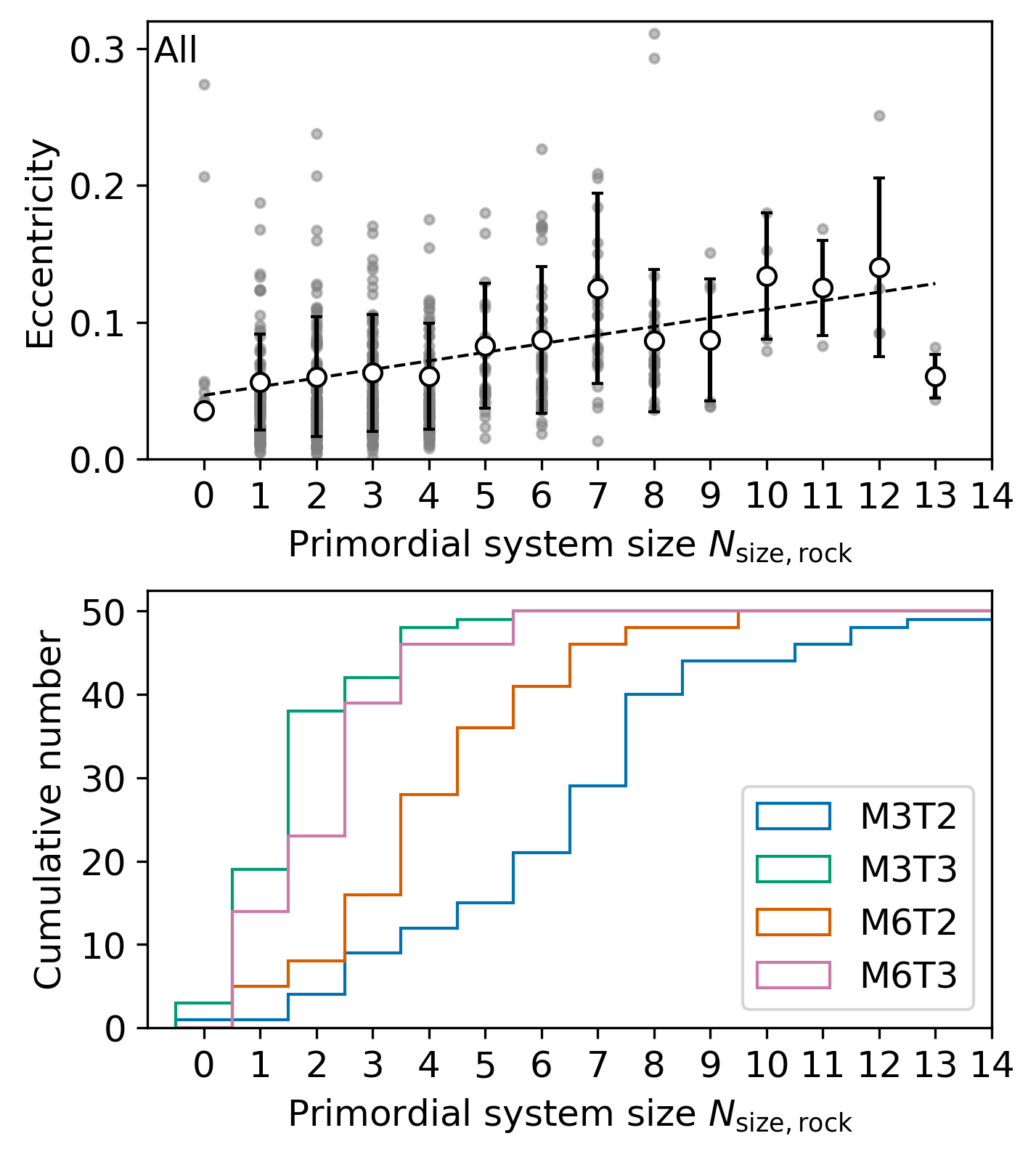}
    \caption{
    {\bf Top:} Eccentricity of rocky planets as a function of the size of the primordial rocky protoplanetary system, $N_\mathrm{size, rock}$. Gray dots represent individual planets, and white markers indicate the mean eccentricity in each bin. The dashed line shows a linear fit to the distribution. 
    {\bf Bottom:} Cumulative distribution of the size of primordial rocky protoplanetary systems, $N_\mathrm{size, rock}$. Each color corresponds to a different scenario.
    }
    \label{fig: Ecc_Mean_Nrock}
\end{figure}

\subsection{\revise{The dynamical origins of elevated eccentricities in the radius valley}}


\revise{Figure~\ref{fig: orbits_evol} shows a representative example in which a rocky planet enters the radius valley after undergoing multiple giant impacts. This outcome is typical across our simulation sets: rocky planets that end up in the radius valley tend to experience a relatively large number of collisions during the giant impact phase, allowing them to grow in size and to have higher eccentricities. As a result, the rocky planets have a positive relation between the planetary radius and the mean eccentricity. On the other hand, the icy protoplanets had fewer giant impacts and formed a rather flat and lower eccentricity distribution. Because of the two distinct populations, the eccentricity distribution peaks between the rocky and icy planet populations. 
}


To understand why some of our formation scenarios produce a population of planets within the radius valley with higher orbital eccentricities while others do not, we first focus on the architecture of planetary systems before disk dispersal. We focus on the size of the system \revise{(planet multiplicity)} at the end of the disk dispersal. We define the ``size'' of a primordial rocky protoplanetary system, $N_\mathrm{size, rock}$, as the number of rocky protoplanets orbiting inside the innermost icy protoplanet (see Appendix for details on the structures of our systems before and after the onset of dynamical instabilities). 

In the top panel of Fig.~\ref{fig: Ecc_Mean_Nrock}, we plot the eccentricity of protoplanets as a function of $N_\mathrm{size, rock}$ and find that eccentricity increases with $N_\mathrm{size, rock}$. For simplicity, we combine the results from all scenarios \revise{in the top panel}; a detailed, scenario-by-scenario analysis for the eccentricity is presented in the next section. The bottom panel of Fig.~\ref{fig: Ecc_Mean_Nrock} shows that in the {\it M3T2} scenario, more than half of the planetary systems have $N_\mathrm{size, rock}$ larger than 6 at the end of the disk phase. \revise{One of the reasons why the  {\it M3T2} scenario produced a peak of eccentricity inside the valley is that it typically produces systems with large $N_\mathrm{size, rock}$. When these systems become dynamically unstable, as shown in Fig.~\ref{fig: orbits_evol}, rocky protoplanets experience several impacts and grow to larger radii and higher eccentricities.} In the other scenarios (see bottom of Figure~\ref{fig: Ecc_Mean_Nrock}), most systems have $N_\mathrm{size, rock}$ smaller than 6, and the dynamical excitation due to instabilities \revise{are not as strong}. 





\begin{figure*}

    \begin{minipage}{0.48\textwidth}  
        \centering
        \includegraphics[width=0.9\linewidth]{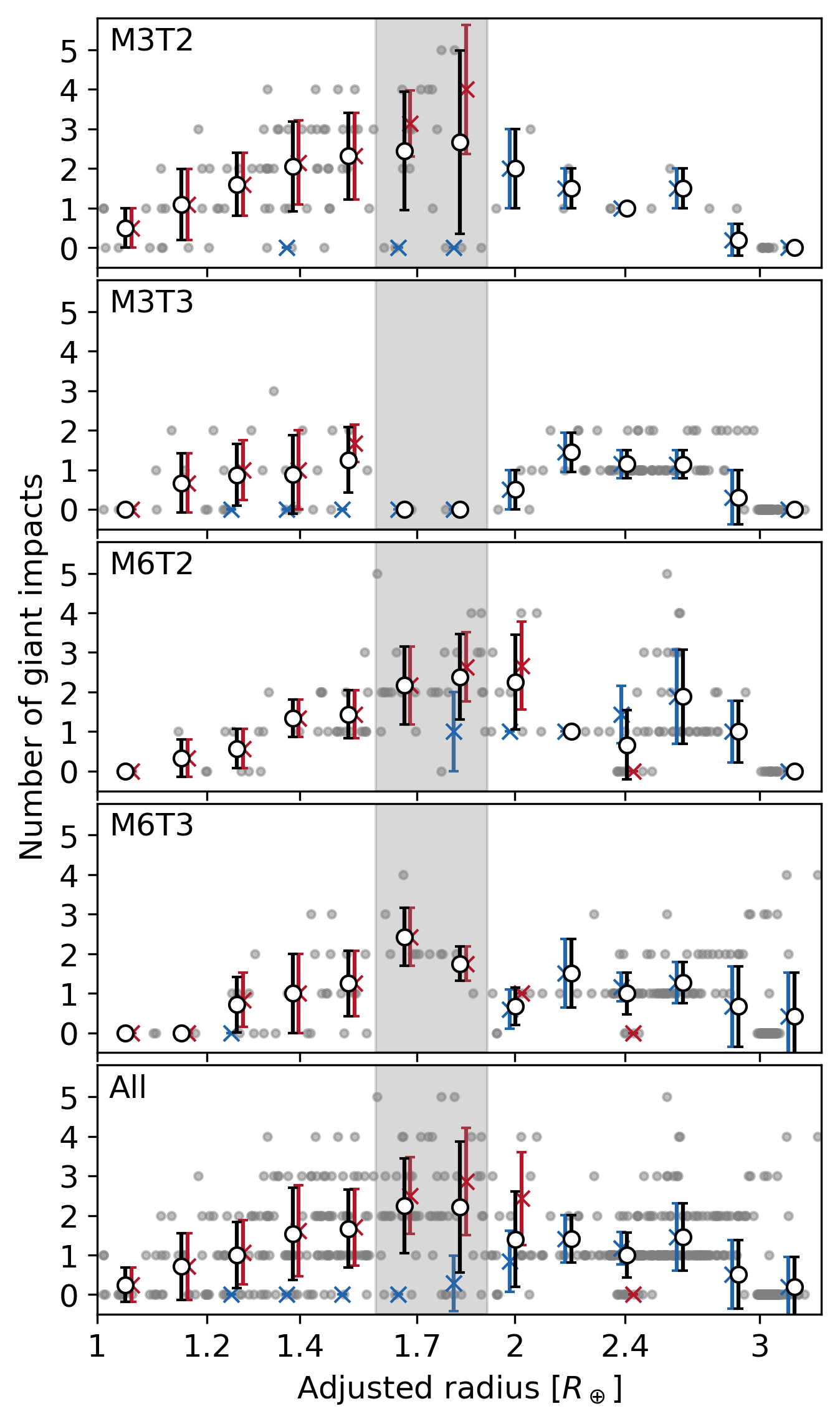}
        \caption{
        Mean number of late giant impacts as a function of the adjusted radius. The white plots show the mean number of late giant impacts. The red and blue plots show those for the rocky and icy planets, respectively.
        }

        \label{fig: Rad_Ncol}
    \end{minipage}
    \hfill
        \centering
    \begin{minipage}{0.48\textwidth}  
        \centering
        \includegraphics[width=0.95\linewidth]{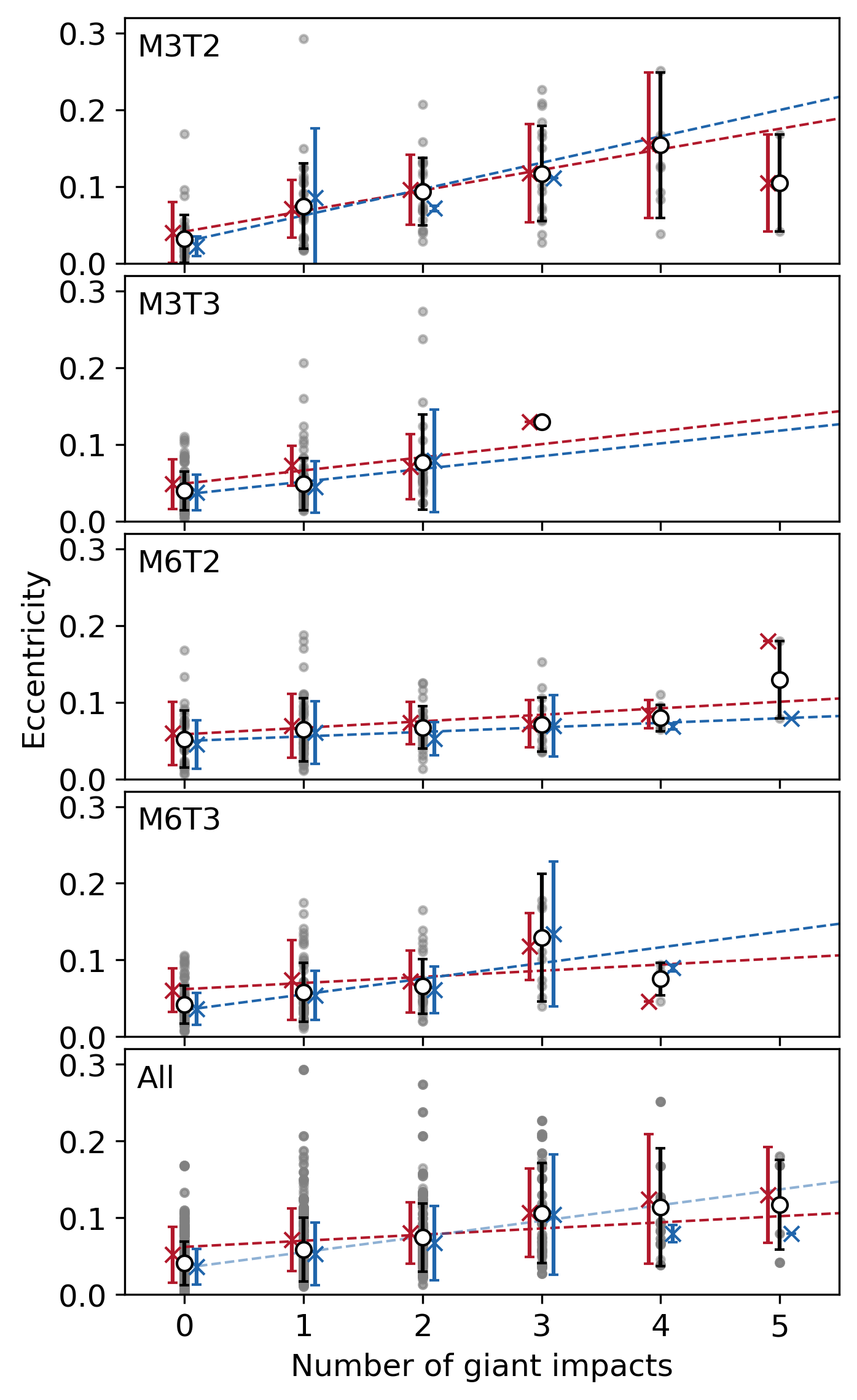} 
        \caption{
        Eccentricity of planets at the end of the simulations as a function of the number of late giant impacts that \revise{each} planet had. The gray plots show the raw data. The white plots are the mean eccentricity, and the error bars show \revise{1-$\sigma$} standard deviation. The dashed lines show the linear fitting to the raw data.
        }

        \label{fig: Ncol_Ecc}
    \end{minipage}
\end{figure*}

\revise{Figure~\ref{fig: Rad_Ncol} shows the correlation between the final adjusted planet radius and its respective number of impacts. It demonstrates that, overall,} planets that entered the radius valley experience, on average, the highest number of late giant impacts after disk dispersal. In particular, rocky planets inside the radius valley in the {\it M3T2} scenario experience an average of two collisions, with some undergoing up to five. In contrast, the {\it M3T3} scenario does not produce any rocky planets inside the radius valley, while rocky planets in the {\it M6T2} and {\it M6T3} scenarios experience, on average, two collisions, with a maximum of three. \revise{Contrary, planets beyond the radius valley are typically icy planets and generally experience relatively fewer impacts. 
}


\revise{In Figure~\ref{fig: Ncol_Ecc}, we plot the mean eccentricity of planets as a function of the number of late giant impacts. Overall, there is a clear positive correlation: planets that undergo more late giant impacts tend to exhibit higher mean eccentricities. This trend is particularly strong in the {\it M3T2} scenario, where planets experiencing three or more impacts reach mean eccentricities nearly twice as large as those with one or no impacts. In comparison, the {\it M3T3}, {\it M6T2}, and {\it M6T3} scenarios also show a positive correlation, but the relationship is weaker. This reflects the comparatively weaker orbital instabilities in these cases.}
Systems with weaker instabilities produce lower planetary eccentricities, preventing the emergence of a clearly distinct, dynamically excited population within the radius valley. Answering why planetary systems produced in {\it M3T2} are subject to stronger instabilities compared to our other models is not trivial, but the distinct dynamical evolutions of the planetary systems shown in Figures~\ref{fig: orbits_evol} and \ref{fig: orbits_evol_M6T2} offer a clue to this question. 

As discussed before, the main difference between those two systems is that the instability shown in Figure~\ref{fig: orbits_evol} involves nearly all protoplanets inside 1~au, resulting in a ``global'' instability event. In contrast, the instability in Figure~\ref{fig: orbits_evol_M6T2} is largely confined to the inner disk and remains virtually decoupled from the outer regions. In the next section, we perform an additional set of simulations to illustrate the main mechanism responsible for enhancing the orbital eccentricities of radius valley planets in our {\it M3T2} simulations, namely the effect of ``external'' perturbers in strengthening dynamical instabilities in the inner system. \revise{We stress that reproducing the eccentricity peak in the radius valley seem to require not only a large number of primordial rocky planets (at the time of disk dispersal), but also the presence of external perturbers  --typically icy planets (e.g. P$>$100 days)-- that enhance dynamical instabilities among planets in the inner system.}




\subsection{Effects of external perturbers: A simplified model}\label{sec: test}

To demonstrate the role of external perturbers, we perform two additional sets of numerical simulations that illustrate how these outer planets can enhance dynamical instabilities in the inner regions of planetary systems. Our initial conditions are inspired by the architecture of the planetary systems produced in the {\it M3T2} scenario at the time of disk dispersal. In our first set of simulations, we distributed 11 rocky protoplanets, each with a mass of $0.5,M_\oplus$, between 0.2 au and 0.5 au, with orbital separations of 10 mutual Hill radii. 
The initial eccentricities are set to zero, and inclinations are randomly set to $\sin i<10^{-3}$.
We carried out 50 simulations, varying the initial orbital angles slightly, and integrated the systems for up to 100~Myr.

\begin{figure}[ht!]
    \plotone{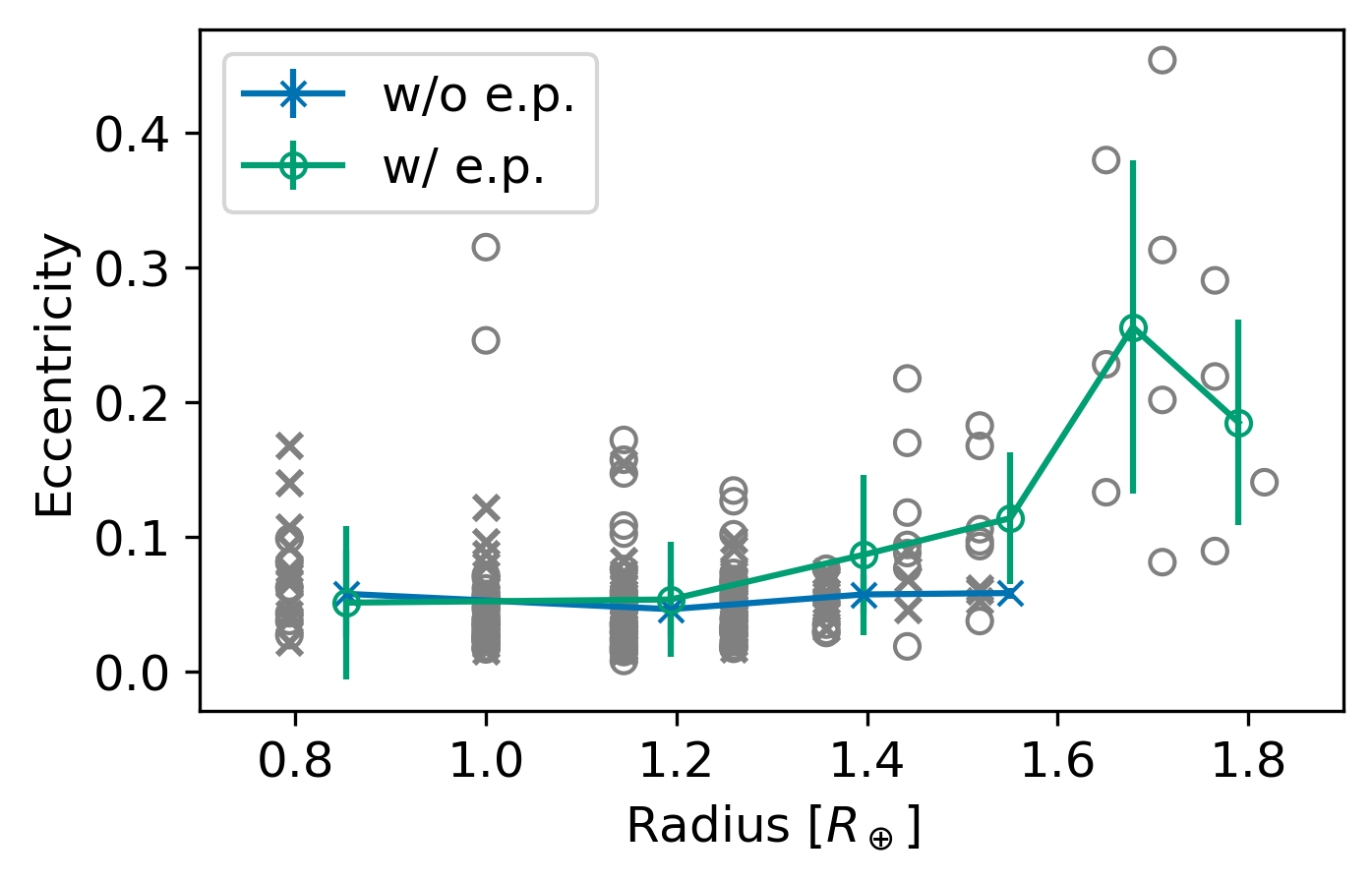}
    \caption{
    Radius v.s. eccentricity of planets obtained in our additional simulations. The blue crosses show the mean eccentricity of planets without external protoplanets (e.p.). The bins have 1 $M_\oplus$ width in the mass coordinate. The green circles show the mean eccentricity of planets with external protoplanets. The gray plots show the raw data.
    }
    \label{fig: Radius_Ecc}
\end{figure}

In all simulations, planetary systems underwent orbital instabilities and giant impacts. The blue line (and crosses) in Figure~\ref{fig: Radius_Ecc} shows the mean eccentricity of the resulting planets orbiting within 100~days after 100 Myr of integration (w/o e.~p. stands for ``without external protoplanets''). In this case, the eccentricity distribution remains broadly ``flat'' as a function of planetary radius. 

Next, we performed an additional set of 50 simulations by taking the same initial systems created with 11 protoplanets between 0.2 au and 0.5 au and adding five external protoplanets, each with a mass of $2M_\oplus$ (with external protoplanets, w/ e.p.). The orbital separation between external protoplanets was set to 10 mutual Hill radii, while 30 mutual Hill radii separated the outermost inner rocky protoplanet and the innermost external protoplanet; external protoplanets are distributed between 0.7 au and 1.4 au.

By the end of the simulations, 60\% of the systems experienced orbital instabilities involving the external protoplanets. As previously discussed, these outer perturbers amplified the dynamical excitation of the inner rocky systems, triggering additional instabilities that facilitated further growth through collisions and elevated eccentricities via interactions with the more massive external bodies. Figure~\ref{fig: Radius_Ecc} shows that, in this case, additional collisions enable planets to grow to radii exceeding 1.6~$R_{\oplus}$ (entering the radius valley), and the mean eccentricity (green line) increases with planetary radius --mirroring the trend observed in the {\it M3T2} scenario.


We conclude that extended planetary systems — characterized by high planet multiplicity at the time of disk dispersal — with external protoplanets can lead to elevated eccentricities around the radius valley. While we do not argue that all planetary systems in the Kepler sample host external perturbers, it is likely that at least a subset of the observed systems was sculpted by such still undetected planets. We also suggest that various parameters related to the system architecture at the time of disk dispersal — such as the protoplanet mass, initial orbital separations, and the number of inner and outer protoplanets — are expected to influence the final eccentricity distribution. Further investigation will be necessary in future studies to fully understand the origin of the observed eccentricity distribution within the broader context of formation models.

\section{Discussion}\label{sec: discussion}

\subsection{How Rocky Disk Mass Shapes Planet Growth and Eccentricity}
\revise{
In our model, rocky and icy protoplanets follow distinct radius distributions, giving rise to a radius valley that separates the two populations. Only rocky planets that experience strong orbital instabilities undergo sufficient growth—via giant impacts—to reach radii within the valley. These same instabilities also excite their orbital eccentricities, naturally producing an overlap between the radius valley and the population of planets with elevated eccentricities. This overlap arises from the combined effects of dynamical excitation and accretion.}

\revise{Since rocky planets in our model accrete most of the available mass from the initial planetesimal disk, their final radii—and thus the radii of eccentric planets—scale with the planetesimal ring mass. In our previous study \citep{Shibata+2025}, we showed that reproducing the observed radius valley requires the inner planetesimal disk to contain typically less than 6 $M_{\oplus}$. If the initial disk mass is too large, planets tend to grow excessively and overshoot the radius valley; conversely, if the disk mass is too low, they fail to reach it at all. This suggests the existence of a ``sweet spot'' in planetesimal disk mass that allows rocky planets to populate the valley.}

\subsection{Caveats}
In this paper, we do not account for observational biases when comparing the final eccentricities of radius valley planets in our simulations with those inferred from observations. Instead, we present a direct, bias-free comparison between our simulation outcomes and the eccentricity distribution derived from observational data. While a more rigorous comparison would be valuable, it would require a significantly larger number of simulations than presented here, as the number of radius valley planets produced in {\it M3T2} is relatively small.

Most eccentric planets within the radius gap in {\it  M3T2} are single planets within $P_\mathrm{orb} \leq 100$ days.  
\revise{As a result, no eccentric planets would be “detected” in the transit observation simulations as part of inner multi-planet systems within $P_\mathrm{orb} \leq 100$ days, assuming observational completeness over this range (but not beyond). This outcome contrasts, in principle, with statistical analyses \citep[e.g.,][]{Gilbert+2025}, which reveal that a subpopulation of eccentric planets within the radius valley exists even in multi-planet systems. Our {\it M3T2} scenario, in fact, appears to overestimate the level of orbital eccentricity for planets in the radius valley (compare Figures~\ref{fig: Rad_Ecc_Obs} and~\ref{fig: Mean_Ecc}), and also produces an excess of single-transit systems. This suggests that a more realistic eccentricity distribution may emerge from a combination of systems with diverse dynamical architectures. For instance, in the {\it M6T2} scenario, we identify several individual systems that host moderately eccentric planets ($e \sim 0.1$) alongside one or two additional planets within $P_\mathrm{orb} = 100$ days. Combining such systems with those from more dynamically violent scenarios could result in a milder overall eccentricity enhancement, better aligning with observations. A detailed exploration of this possibility is left for future work. Moving forward, it will be important to broaden the range of initial conditions in the formation model, incorporate observational biases, and generate synthetic transit populations to test whether a mixed population of dynamically quiet and unstable systems can fully account for the observed eccentricity structure across the radius valley.}

\subsection{Eccentricity enhancement of icy planets}

\revise{
The inferred eccentricity distribution of exoplanets by \citet{Gilbert+2025} suggests that the mean eccentricity increases from $2R_{\oplus}$ to about $3.5R_{\oplus}$. This trend implies that planets larger than $2 R_{\oplus}$, typically icy planets in our model, tend to exhibit higher orbital eccentricities. One possible explanation is that orbital instabilities among icy protoplanets, analogous to those affecting rocky planets, may drive this eccentricity enhancement. However, our numerical results do not reproduce this trend, likely because our model underestimates the number of icy planets in close-in orbits. 
}

\revise{
As shown by the architectures of planetary systems in the {\it M3T2} and {\it M3T3} scenarios (see Appendix), longer-lived disks tend to produce a larger number of icy planets in close-in orbits. In our simulations, we considered only disk lifetimes of $\tau_\mathrm{disk,life} = 2$ and 3 Myr. However, observations of protoplanetary disks indicate a broader distribution of lifetimes \citep{Richert+2018}. Longer disk lifetimes allow more time for icy protoplanets to migrate inward from beyond the snow line, thereby increasing the number of icy planets in the inner system. Future studies should investigate how extending the disk lifetime affects the formation and inward transport of icy protoplanets, and how these changes influence the dynamics of late-stage instabilities, the resulting planet compositions, and the structure of the radius valley.
}

\section{Summary}\label{sec: conclusion}

We revisited recent simulations of the breaking the chains model for the formation of super-Earths and mini-Neptunes, as developed in \citet{Izidoro+2022,Shibata+2025}. These models broadly reproduce key observational properties of close-in planets, including period ratios, size ratios, multiplicities, and the bimodal size distribution known as the radius valley. Building on this framework, we investigated whether the elevated eccentricities inferred for planets within the radius valley \citep{Gilbert+2025} can be explained by late-stage dynamical instabilities and giant impacts among rocky planets.

Our results suggest that the radius valley—and the planets residing within it—is shaped by both post-disk dynamical instabilities and a fundamental compositional dichotomy between rocky super-Earths and water-rich mini-Neptunes. We find that systems with high rocky protoplanet multiplicity and the presence of external perturbers are more likely to undergo violent dynamical instabilities after disk dispersal. These instabilities, amplified by gravitational interactions with the outer perturbers, lead to a greater number of giant impacts among inner rocky planets. The resulting collisions produce dynamically excited, intermediate-size rocky planets that tend to populate the radius valley. While the precise level of eccentricity excitation depends on the system’s initial architecture, our simulations reproduce the key trends seen in current observational data, supporting a dynamical origin for the elevated eccentricities of valley planets reported in \citet{Gilbert+2025}.

This study reinforces the view that the radius valley is not merely a compositional divide \citep{Owen+2017, Zeng+2019}, but also a dynamical one -- reflecting the interplay of dynamical instabilities and giant impacts.

\begin{acknowledgments}
We are very thankful to Bertram Bitsch for valuable discussions. This work was supported in part by the Big-Data Private-Cloud Research Cyberinfrastructure MRI-award funded by NSF under grant CNS-1338099 and by Rice University's Center for Research Computing (CRC). Part of the numerical computations were carried out on PC cluster at the Center for Computational Astrophysics, National Astronomical Observatory of Japan. 
\end{acknowledgments}

\appendix




\section{Planetary system structures}\label{app: appendix2}

\begin{figure}[ht!]
    \centering
    \includegraphics[width=\linewidth]{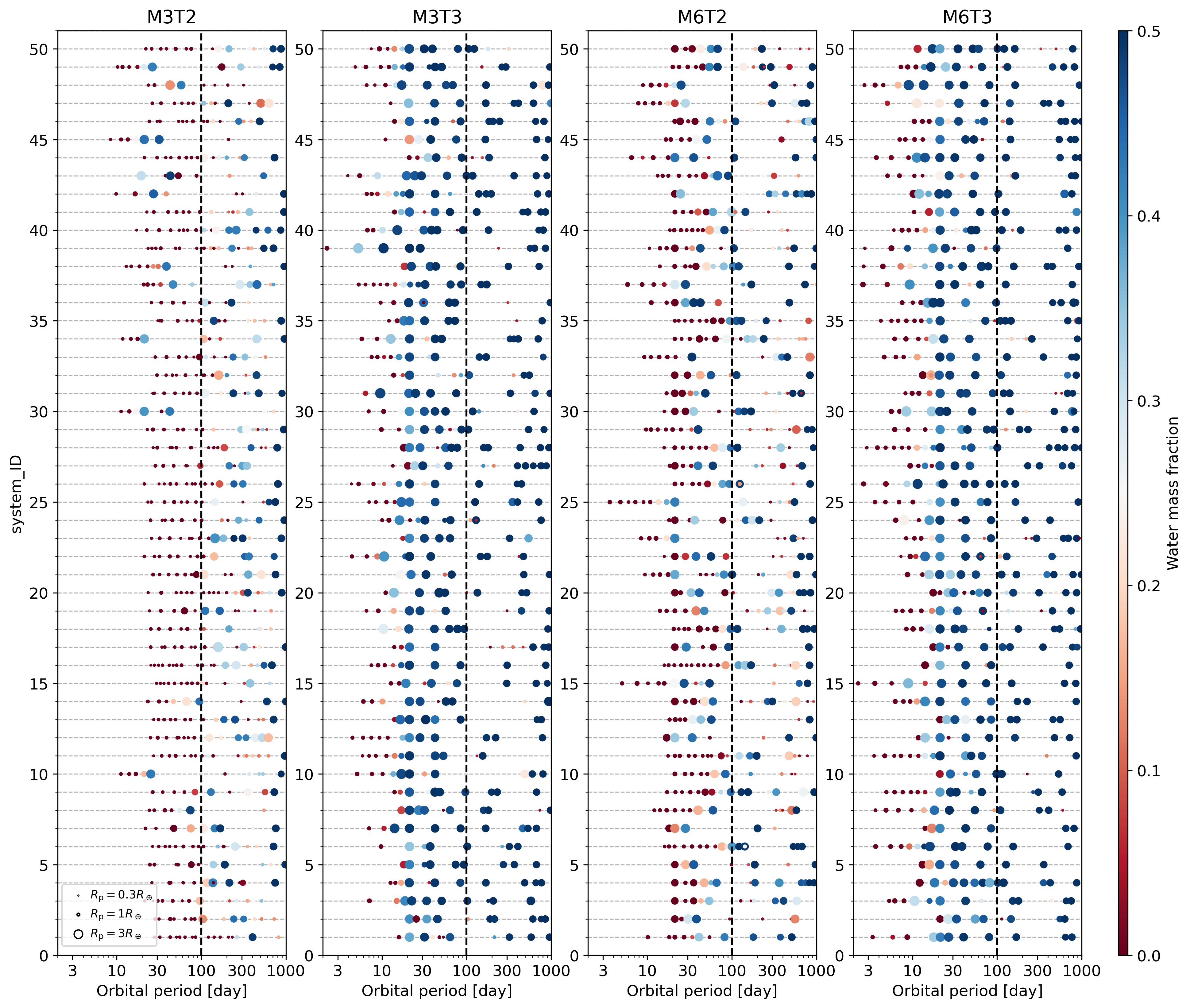}
    \caption{
    The structure of planetary systems at the disk dissipation time.
    }
    \label{fig: Porb_systems}
\end{figure}

\begin{figure}[ht!]
    \centering
    \includegraphics[width=\linewidth]{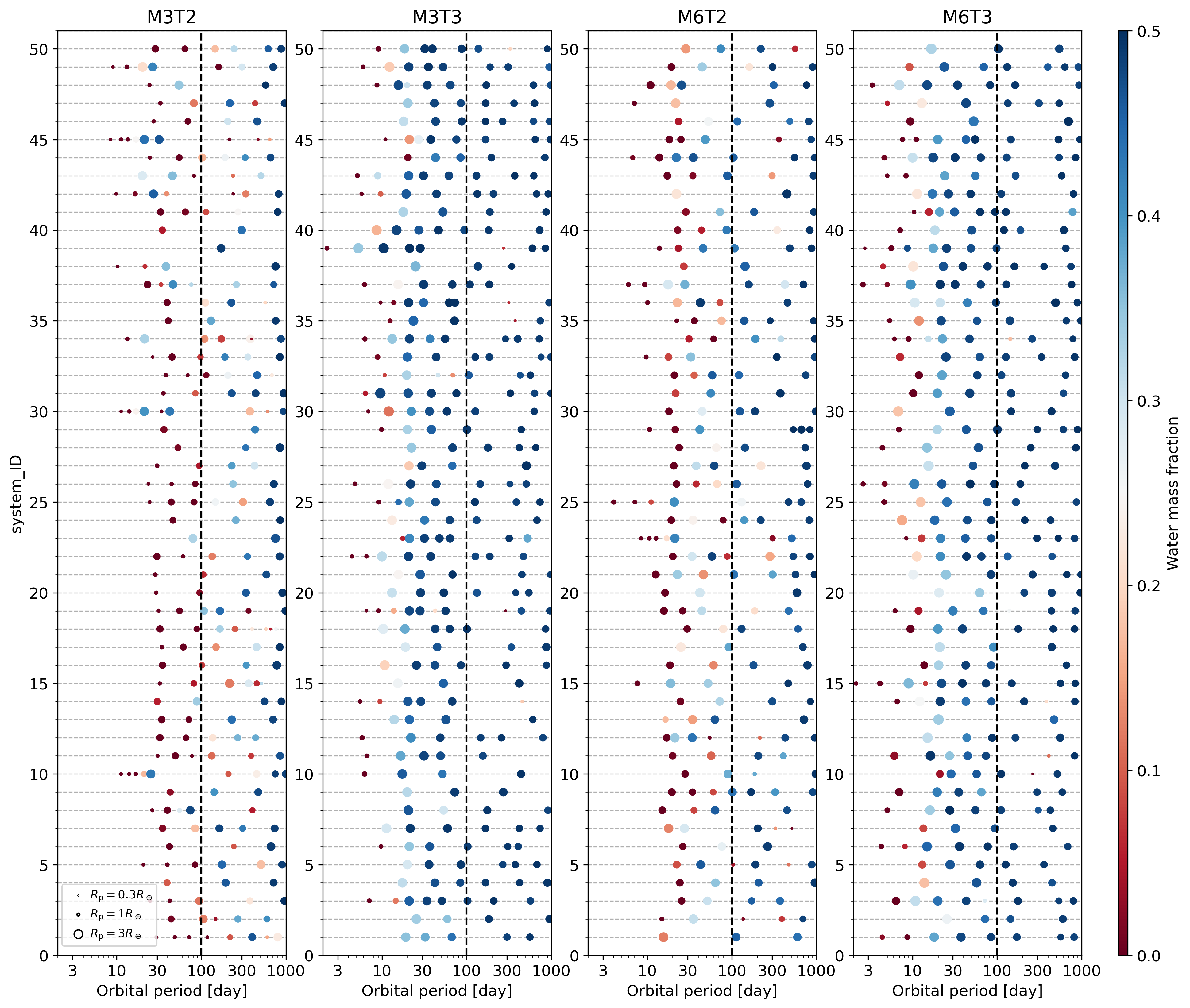}
    \caption{
    The structure of planetary systems at the end of the simulations.
    }
    \label{fig: Porb_systems_end}
\end{figure}

We show the architecture of planetary systems at the disk dissipation time $t=\tau_\mathrm{disk, life}$ and at the end of the simulations $t=50$ Myrs in Fig.~\ref{fig: Porb_systems} and \ref{fig: Porb_systems_end}.


\bibliography{ref}{}
\bibliographystyle{aasjournal}



\end{document}